\documentclass[prb,preprint]{revtex4-1}

\usepackage{amsmath}  
\usepackage{amssymb}
\usepackage{amsfonts} 
\usepackage{graphicx} 
\usepackage{xcolor}

\newcommand{\simlt}
{\lower.5ex\hbox{\ltsima}}
\newcommand{\simgt}
{\lower.5ex\hbox{\gtsima}}

\begin{document}
\def\bigint{{\displaystyle\int}}
\def\simlt{\stackrel{<}{{}_\sim}}
\def\simgt{\stackrel{>}{{}_\sim}}

\title{Why physical understanding should precede the mathematical formalism - conditional quantum probabilities as a case-study}

\author{Yakir Aharonov}
\email{yakir@post.tau.ac.il} 
\affiliation{Institute for Quantum Studies, Chapman University, Orange, CA 92866, USA}
\affiliation{Schmid College of Science and Technology, Chapman University, Orange, CA 92866, USA}
\affiliation{School of Physics and Astronomy, Tel Aviv University, Tel Aviv 6997801, Israel}
\author{Eliahu Cohen}
\email{eliahu.cohen@biu.ac.il} 
\affiliation{Faculty of Engineering and the Institute of Nanotechnology and Advanced Materials, Bar Ilan University,
Ramat Gan 5290002, Israel}

\author{David H. Oaknin}
\email{d1306av@gmail.com} 
\affiliation{Rafael Advanced Defense Systems Ltd., Haifa 61532, Israel}

\date{\today}

\begin{abstract}
Conditional probabilities in quantum systems which have both initial and final boundary conditions are commonly evaluated using the Aharonov-Bergmann-Lebowitz rule. In this short note we present a seemingly disturbing paradox that appears when applying the rule to systems with slightly broken degeneracies. In these cases we encounter a singular limit -- the probability ``jumps'' when going from perfect degeneracy to negligibly broken one. We trace the origin of the paradox and solve it from both traditional and modern perspectives in order to highlight the physics behind it: the necessity to take into account the finite resolution of the measuring device. As a practical example, we study the application of the rule to the Zeeman effect. The analysis presented here may stress the general need to first consider the governing physical principles before heading to the mathematical formalism, in particular when exploring puzzling quantum phenomena.
\end{abstract}

\maketitle

\section{Introduction}

It is widely accepted that any physical theory must be formulated, once its fundamental principles have been understood and established, as a sound mathematical model in which the properties of all entities involved and their relationships are precisely described. Only then can the theory and its predictions be accurately confronted with the experimental evidence.

In some cases, the confrontation of the theoretical predictions with the data collected from experiments led to the conclusion that the tested theory is only an approximation valid within a limited range of applications of a more general theory. For example, Galileo's transformation law defined with respect to different inertial frames is only an approximation of the more general Lorentz transformation law, classical electromagnetism is only an approximation of quantum electrodynamics and Newton's theory of gravitation is only an approximation of Einstein's general relativity.

In other cases, the fundamental principles of the theory and its range of applicability are well established but the predictions obtained from them involve certain approximations whose range of validity may be limited. Hence, such predictions will be valid as long as the approximations involved are valid. In such cases, it is of utmost importance not to forget the physical understanding of the assumptions and approximations involved in the mathematical formalism that led to the predictions of interest. Otherwise, a misplaced use of the formalism may lead, as we shall show, to apparent paradoxes. In such cases, we believe that one should return to the fundamental physical principles lying at the heart of the problem and solve it carefully using an appropriate mathematical formalism that fits these principles. This perspective has been implicitly adopted in our previous works \cite{AR,P1,P2,P3,P4}, but here we wish to make it very explicit and further emphasize its importance when exploring quantum phenomena, especially when relying on a purely informational perspective  \cite{Fuchs1,Fuchs2,Bub1,Bub2,Fuchs3}.


We shall base this perspective on a particular example concerned with conditional probabilities in quantum mechanics. This topic has previously been a source of puzzles and debates (see for example \cite{ABL1,ABL2,ABL3,ABL4,ABL5,ABL6,ABL7,ABL8}), but here we point out a new, at first perplexing paradox, which gives the impression of a singular limit \cite{Berry}. We shall examine an apparent peculiarity related to the calculation of probabilities in slightly degenerate quantum systems which have both initial and final boundary conditions. These conditional probabilities are commonly evaluated using the Aharonov-Bergmann-Lebowitz (ABL) rule \cite{ABL}, which will now be examined from both traditional and modern perspectives to highlight the physics behind it, especially when closely separated energy levels are involved.

To this aim we will resort to the modern notion of quantum measurements described by a set of positive-operator valued measure~(POVM) elements, which generalizes the more traditional notion of projective measurements. Projective measurements are described by the set of projectors $\left\{{\hat P}_k\right\}_{k \in K}$ onto the eigenspaces of each of the eigenvalues $\left\{c_k\right\}_{k \in K}$ of the measured quantum observable ${\hat C}$, so that

\begin{equation}
{\hat C} = \sum_{k \in K} \ c_k \cdot {\hat P}_k,
\end{equation}
while
\begin{equation}
\sum_{k \in K} \ {\hat P}_k = \mathbb{I},  \hspace{0.8in} {\hat P}^{\dagger}_k = {\hat P}_k, \hspace{0.8in} {\hat P}_k \cdot {\hat P}_{k'} = \delta_{k, k'} {\hat {\mathbb{I}}}.
\end{equation}
Hence, the index $k$ labels the possible different outcomes $c_k$ of the projective measurement. The Born rule provides the probability for each of these outcomes to occur
\begin{equation}
p_k = \langle \Psi|{\hat P}_k|\Psi\rangle,
\end{equation}
when the quantum system is prepared in a state $|\Psi\rangle$. It is then straightforward to test that
\begin{equation}
\sum_{k \in K} p_k = \langle \Psi|\sum_{k \in K} {\hat P}_k|\Psi\rangle = \langle \Psi|{\hat {\mathbb{I}}}|\Psi\rangle = 1.
\end{equation}
Upon measurement, the quantum state of the system gets projected onto the corresponding eigenspace:
\begin{equation}
|\Psi\rangle \longrightarrow \frac{1}{\sqrt{p_k}} {\hat P}_k|\Psi\rangle.
\end{equation}
A generalized measurement is described by a family $\left\{{\hat {\cal E}}_r \equiv {\hat {\cal F}}^{\dagger}_r {\hat {\cal F}}_r \right\}_{r \in R}$ of positive-valued operators on the Hilbert space of the considered quantum system such that:
\begin{equation}
\sum_{r \in R} \ {\hat {\cal E}}_r = {\hat {\mathbb{I}}},
\end{equation}
where $R$ is the family of possible outcomes. Each one of them occurs with probability
\begin{equation}
p_r = \langle \Psi|{\hat {\cal E}}_r|\Psi\rangle = \langle \Psi|{\hat {\cal F}}^{\dagger}_r {\hat {\cal F}}_r|\Psi\rangle = \left|{\hat {\cal F}}_r|\Psi\rangle\right|^2 \ge 0,
\end{equation}
when the quantum system is prepared in the state $|\Psi\rangle$,  which generalizes the standard Born rule for projective measurements. As before, it is straightforward to test that
\begin{equation}
\sum_{r \in R} p_r = \sum_{r \in R} \langle \Psi|{\hat {\cal E}}_r|\Psi\rangle = \langle \Psi|\sum_{r \in R} {\hat {\cal E}}_r|\Psi\rangle = \langle \Psi|{\hat {\mathbb{I}}}|\Psi\rangle = 1.
\end{equation}\
Upon measurement, the quantum state of the system gets projected as:
\begin{equation}
|\Psi\rangle \longrightarrow \frac{1}{\sqrt{p_r}} {\hat {\cal F}}_k|\Psi\rangle,
\end{equation}
which is not necessarily an eigenstate of the measured observable. The formalism of generalized measurements is needed, as we shall show below, when the projective measurement actually involves a system larger than the considered subsystem of interest. This larger system may include, for example, the measuring device, which is indeed the situation below.

\section{A case-study concerning conditional quantum probabilities}

In the standard formalism of quantum mechanics the most complete description of a closed system at a time $t_i$ is given by a state vector $|\Psi(t_i)\rangle$, which is determined based on the outcomes of a complete set of compatible measurements performed on it until the given time. If the system is then left isolated, its state is determined at any later time $t \ge t_i$ by the Schr\"odinger equation

\begin{equation}
\label{forward}
|\Psi(t)\rangle = U(t,t_i) |\Psi(t_i)\rangle,
\end{equation}
where $U(t,t_i)=e^{-i {\hat H} (t-t_i)}$ and ${\hat H}$ is the time-independent Hamiltonian of the system.

The description (\ref{forward}) holds over the time interval $t_i \le t < t_f$ until $t_f$ at which a new set of compatible measurements is performed on the system, whose outcomes allow to redefine the state of the system via $|\Phi(t_f)\rangle$. Similarly to the above description we could say that during the time interval $t_i < t \le t_f$ the quantum system is described by the state vector given by

\begin{equation}
\label{backward}
|\Phi(t)\rangle = U(t_f,t)^{\dagger} |\Phi(t_f)\rangle.
\end{equation}

Given these initial and final conditions at times $t_i$ and $t_f$ we can ask about the probability to obtain a given outcome $c_k$ in a projective measurement of a physical observable ${\hat C}$ at any intermediate time $t \in (t_i, t_f)$, where $c_k \in \left\{c_{k'}\right\}_{k' \in K}$ is one of the eigenvalues of the measured observable. This question was explored in \cite{ABL,ABL1} and it was found that

\begin{equation}
\label{conditional}
p({\hat C}=c_k) = \frac{\left|\langle \Phi(t)|{\hat P}_k|\Psi(t)\rangle \right|^2}{\sum_{k' \in K} \left|\langle \Phi(t)|{\hat P}_{k'}|\Psi(t)\rangle \right|^2},
\end{equation}
where ${\hat P}_k$ denotes the projector onto the eigenspace of the corresponding eigenvalue $c_k$ for every $k \in K$. It can be readily verified that $p({\hat C}=c_k) \in [0, 1]$ and $\sum_{k' \in K} p({\hat C}=c_{k'}) = 1$, hence the relation (\ref{conditional}) can be thought of as a conditional probability, given the initial and final states of the system. The relation (\ref{conditional}) is important in that it restores time-symmetry, even when collapse is involved in the description of quantum systems.

In this note we focus on the particular case in which the considered physical observable takes the form ${\hat C} = {\hat C}_0 + \epsilon {\hat C}_1$, where $\epsilon \in \mathbb{R}$ is some small real number and the operators ${\hat C}_0$ and ${\hat C}_1$ do commute $\left[{\hat C}_0,{\hat C}_1\right]=0$. Besides, we shall assume that ${\hat C}_0$ has a degenerate spectrum, while the perturbation ${\hat C}_1$ breaks some or all these degeneracies. It can be readily seen that the conditional probability as defined in (\ref{conditional}) shows a discontinuity at $\epsilon=0$.

As an example, we consider a three dimensional system and denote by $\left\{|\alpha\rangle, |\beta\rangle, |\gamma\rangle\right\}$ an orthonormal basis. We define the linear operators ${\hat C}_0$ and ${\hat C}_1$ through the relationships:

\begin{eqnarray}
{\hat C}_0|\alpha\rangle  = -|\alpha\rangle, \hspace{0.5in}
{\hat C}_0|\beta\rangle = +|\beta\rangle, \hspace{0.5in}
{\hat C}_0|\gamma\rangle = +|\gamma\rangle,
\end{eqnarray}
and

\begin{eqnarray}
{\hat C}_1|\alpha\rangle  = -|\alpha\rangle, \hspace{0.5in}
{\hat C}_1|\beta\rangle = +|\beta\rangle, \hspace{0.5in}
{\hat C}_1|\gamma\rangle = +2|\gamma\rangle.
\end{eqnarray}
Hence, according to (\ref{conditional}) we have

\begin{eqnarray}
\nonumber
p({\hat C}(\epsilon = 0)=-1) \ \ \ \ \   & = & \frac{\left|\langle \Phi(t)|\alpha\rangle \cdot \langle\alpha|\Psi(t)\rangle \right|^2}{\left|\langle \Phi(t)|\alpha\rangle \cdot \langle\alpha|\Psi(t)\rangle \right|^2 + \left|\langle \Phi(t)|\beta\rangle \langle\beta|\Psi(t)\rangle + \langle \Phi(t)|\gamma\rangle \langle\gamma|\Psi(t)\rangle \right|^2},
\nonumber
\\
\nonumber
\\
\nonumber
p({\hat C}(\epsilon \neq 0)=-1-\epsilon) & = & \frac{\left|\langle \Phi(t)|\alpha\rangle \cdot \langle\alpha|\Psi(t)\rangle \right|^2}{\left|\langle \Phi(t)|\alpha\rangle \cdot \langle\alpha|\Psi(t)\rangle \right|^2 + \left|\langle \Phi(t)|\beta\rangle \cdot \langle\beta|\Psi(t)\rangle \right|^2 + \left|\langle \Phi(t)|\gamma\rangle \cdot \langle\gamma|\Psi(t)\rangle \right|^2}. \\
\end{eqnarray}
In the denominator of the expression for $p({\hat C}(\epsilon = 0)=-1)$ appears the term \newline
$\left|\langle \Phi(t)|\beta\rangle \langle\beta|\Psi(t)\rangle + \langle \Phi(t)|\gamma\rangle \langle\gamma|\Psi(t)\rangle \right|^2$, since the eigenvalue $C=1$ is degenerate when $\epsilon=0$. On the other hand, when $\epsilon \neq 0$ the degeneracy is broken into two different non-degenerate eigenvalues: $C = 1 + \epsilon$ and $C = 1 + 2 \epsilon$ and, therefore, in the denominator of the expression for $p({\hat C}(\epsilon \neq 0)=-1-\epsilon)$ appears the term $\left|\langle \Phi(t)|\beta\rangle \cdot \langle\beta|\Psi(t)\rangle \right|^2 + \left|\langle \Phi(t)|\gamma\rangle \cdot \langle\gamma|\Psi(t)\rangle \right|^2$, which is not necessarily equal to the former. Indeed, it is necessarily equal or larger, so that

\begin{equation}
p({\hat C}(\epsilon \neq 0)=-1-\epsilon) \ \ \le \ \ p({\hat C}(\epsilon = 0)=-1).
\end{equation}
 \

At first sight, the discontinuity at $\epsilon=0$ might be surprising and even disturbing, since it would imply that the hypothetical measurement at intermediate time would be able to detect, through a finite jump in the frequencies of its possible outcomes, any $\epsilon \neq 0$ no matter how small it could be. For example, in the simple example that we have just discussed the measurement seems to be able to detect a difference between the two largest eigenvalues as small as, say, $\epsilon=10^{-100}$ or even smaller, which is of course physically unreasonable.

However, as implied above, one has to take into account the physical principles involved in order to understand this apparent conundrum. It turns out that the uncertainty of the measuring pointers has to be taken into account in order to understand the aforementioned discontinuity. Namely, the outcome of a measurement corresponds to a shift in the position of a pointer, whose wavefunction always has, in practice, a non-zero width. The ABL rule as stated in (\ref{conditional}) is valid for projective measurements for which, by definition, the uncertainty of the measuring pointers is negligible compared to the gap between the different eigenvalues of the measured observable. The finite precision of the measuring pointer must obviously be taken into account when it is larger or comparable to the gap between different eigenvalues, since then they become practically indistinguishable and the system should be thought{\color{magenta}s} of as being effectively degenerate.

In order to take into account the uncertainty in the pointer's state, we can describe the measurement at time $t$ by a set $\left\{{\hat {\cal E}}_r \equiv {\hat {\cal F}}^{\dagger}_r {\hat {\cal F}}_r\right\}_{r \in R}$ of positive-operator valued measure (POVM) elements

\begin{equation}
\sum_{r \in R} {\hat {\cal E}}_r = \hat{\mathbb{I}},
\end{equation}
rather than as a projective measurement. The ABL rule can then be readily generalized as follows

\begin{equation}
\label{conditional1}
p({\hat {\cal F}}=f_r) = \frac{\left|\langle \Phi(t)|{\hat {\cal F}}_r^{\dagger} {\hat {\cal F}}_r|\Psi(t)\rangle \right|^2}{\sum_{r' \in R} \left|\langle \Phi(t)|{\hat {\cal F}}_{r'}^{\dagger} {\hat {\cal F}}_{r'}|\Psi(t)\rangle \right|^2}.
\end{equation}
Within this generalized framework we can smoothly connect the case of a non-degenerate system in which the precision of the pointer is much larger than the gap between its energy eigenvalues to the case in which the width of the pointer is much smaller than the gap between eigenvalues.
\\
For example, a generalized measurement at time $t$ could be defined as follows:

\begin{equation}
{\cal F}_{\alpha} = |\alpha\rangle \langle \alpha|, \hspace{0.2in} {\cal F}_{\beta} = \frac{1}{\sqrt{1 + {\cal \zeta}^2}}\left(|\beta\rangle\langle\beta| + \zeta |\gamma\rangle \langle\gamma|\right),  \hspace{0.2in} {\cal F}_{\gamma} =   \frac{1}{\sqrt{1 + {\cal \zeta}^2}}\left(\zeta |\beta\rangle \langle\beta| + |\gamma\rangle\langle\gamma|\right),
\end{equation}
where $\zeta \equiv e^{-\frac{\epsilon^2}{4\Delta^2}}$ and $\Delta$ is the resolution of the measurement, i.e. the width of the Gaussian pointer. It can be readily checked that

\begin{eqnarray*}
{\cal F}^{\dagger}_{\alpha} {\cal F}_{\alpha} & = & |\alpha\rangle \langle \alpha|, \\
{\cal F}^{\dagger}_{\beta} {\cal F}_{\beta} & = & \frac{1}{1 + {\cal \zeta}^2}\left(|\beta\rangle\langle\beta| + \zeta^2 |\gamma\rangle \langle\gamma|\right), \\
{\cal F}^{\dagger}_{\gamma} {\cal F}_{\gamma} & = & \frac{1}{1 + {\cal \zeta}^2}\left(\zeta^2 |\beta\rangle \langle\beta| + |\gamma\rangle\langle\gamma|\right),
\end{eqnarray*}
 since $\langle\beta|\gamma\rangle = 0$. Hence,

\begin{equation}
{\cal F}^{\dagger}_{\alpha} {\cal F}_{\alpha} + {\cal F}^{\dagger}_{\beta} {\cal F}_{\beta} + {\cal F}^{\dagger}_{\gamma} {\cal F}_{\gamma} = \hat {\mathbb{I}},
\end{equation}
as required.

Moreover, it is also straightforward to notice that in the limit of $\epsilon \gg \Delta$, we have $\zeta \simeq 0$ and therefore,

\begin{equation}
{\cal F}_{\alpha} = |\alpha\rangle \langle \alpha|, \hspace{0.2in} {\cal F}_{\beta} \simeq |\beta\rangle\langle\beta|,  \hspace{0.2in} {\cal F}_{\gamma} \simeq |\gamma\rangle \langle\gamma|.
\end{equation}
This limit describes a measurement whose resolution is able to distinguish between the eigenvalues $C = 1+\epsilon$ and $C = 1+2\epsilon$ of the observable ${\hat C}$ and, indeed, in this limit equation (\ref{conditional1}) reproduces the ABL rule for the non-degenerate case $\epsilon \neq 0$.

On the other hand, in the limit of $\epsilon \ll \Delta$, we have $\zeta \simeq 1$ and, therefore,

\begin{equation}
{\cal F}_{\alpha} = |\alpha\rangle \langle \alpha|, \hspace{0.2in} {\cal F}_{\beta} \simeq {\cal F}_{\gamma} \simeq \frac{1}{\sqrt{2}}\left(|\beta\rangle\langle\beta| + |\gamma\rangle \langle\gamma|\right).
\end{equation}
This limit describes a measurement whose resolution is not able to distinguish between the eigenvalues $C = 1 + \epsilon$ and $C = 1 + 2\epsilon$ associated to the eigenstates $|\beta\rangle$ and $|\gamma\rangle$, respectively. In this limit, equation (\ref{conditional1}) reproduces the ABL rule for the degenerate case $\epsilon = 0$.

Equation (\ref{conditional1}), which trivially generalizes the original ABL formula for projective measurements (\ref{conditional}), can be obtained as follows:

In the standard von-Neumann description of a projective measurement the probed system and the measuring device (pointer) are prepared in a separable state and then interact for a short time interval, after which the state of the coupled system is described by the unitary transformation

\begin{eqnarray*}
e^{i \theta {\hat C} \otimes {\hat P}} = e^{i \theta \sum_{k} c_k |k\rangle\langle k| \otimes {\hat P}} = \prod_{k} e^{i \theta  c_k |k\rangle\langle k| \otimes {\hat P}} = \prod_{k} \left(\hat {\mathbb{I}}+ |k\rangle\langle k| \otimes \left(e^{ i \theta  c_k {\hat P}} - \hat {\mathbb{I}}\right)\right) = \\
= \hat {\mathbb{I}}+ \sum_{k} |k\rangle\langle k| \otimes \left(e^{ i \theta  c_k {\hat P}} - \hat {\mathbb{I}}\right)
= \sum_{k} |k\rangle\langle k| \otimes e^{ i \theta  c_k {\hat P}} = \sum_{k \in K} {\hat P}_k \otimes e^{ i \theta  c_k {\hat P}},
\end{eqnarray*}
where as before ${\hat C}$ is the tested observable, $|k\rangle$ and $c_k$ are its eigenvectors and corresponding eigenvalues, ${\hat P}$ is the momentum operator of the pointer and $\theta$ describes the strength of the interaction. If before the interaction the pointer is prepared in an eigenstate $|0\rangle$ of its position operator, the transformation $e^{ i \theta  c_k {\hat P}} |0\rangle = |x_k\rangle$ describes a shift in the position of the pointer by an amount $x_k = \theta c_k$. Hence,

\begin{equation}
e^{i \theta {\hat C} \otimes {\hat P}} |0\rangle = \sum_{k \in K} {\hat P}_k \otimes e^{ i \theta  c_k {\hat P}} |0\rangle =
\sum_{k \in K} {\hat P}_k \otimes |x_k\rangle.
\end{equation}
Therefore, a shift $x_k$ in the position of the pointer at the end of the measurement would imply that the state of the measured system have been projected by the projector ${\hat P}_k$ onto the eigenspace of the corresponding eigenvalue $c_k$. This situation would correspond to the original ABL rule (\ref{conditional}). Let us note, for the sake of clarity in the discussion that follows, that for any projector ${\hat P}^{\dagger}_k {\hat P}_k = {\hat P}_k^2 = {\hat P}_k$.

Similarly, for any other orthonormal basis of states of the pointer $\left\{|y_r\rangle\right\}_{r \in R}$, such that its position eigenstates can be written as $|x_k\rangle = \sum_{r \in R} \rho_{k,r} |y_r\rangle$ with $\sum_{r \in R} \left|\rho_{k,r}\right|^2 = 1$, we have

\begin{equation}
e^{i \theta {\hat C} \otimes {\hat P}} |0\rangle = \sum_{k \in K} {\hat P}_k \otimes |x_k\rangle = \sum_{k \in K} {\hat P}_k \otimes
\left(\sum_{r \in R} \rho_{k,r} |y_r\rangle\right) = \sum_{r \in R} \left(\sum_{k \in K} \rho_{k,r} {\hat P}_k\right) \otimes |y_r\rangle.
\end{equation}
In this case an outcome $y_r$ at the end of the measurement would imply that the state of the measured system has been transformed by the operator ${\cal F}_r = \sum_{k \in K} \rho_{k,r} {\hat P}_k$, which would lead to the generalized ABL rule (\ref{conditional1}). It is
straightforward to test that

\begin{equation}
\sum_{r \in R} {\cal F}^{\dagger}_r {\cal F}_r = \sum_{k \in K} \sum_{k' \in K} \sum_{r \in R} \rho_{k',r}^* \rho_{k,r} {\hat P}_{k'} {\hat P}_k = \sum_{k \in K} \sum_{r \in R} |\rho_{k,r}|^2 {\hat P}_k = \sum_{k \in K} {\hat P}_k = \hat {\mathbb{I}}.
\end{equation}

\section{An example: The Zeeman effect in the hydrogen atom}

As a simple and deductive example we consider the electronic degrees of freedom in a hydrogen atom. For the sake of simplicity we ignore the electron's spin and neglect relativistic effects. In such approximation the energy eigenstates of the electronic Hamiltonian ${\hat C}$ (although the Hamiltonian is usually denoted by $\hat{H}$ we keep using the above notation for the measured observable due to pedagogic reasons) can be labelled by three quantum numbers as $\left|n, \  l, \ m \right.\rangle$, where $n=1,2,3,....$ is the principal quantum number, $l=0,...,n-1$ labels the orbital angular momentum of the electron and $m=-l,...,l$ is the projection of the orbital angular momentum of the electron along the $Z$ axis.

In the absence of external magnetic fields the corresponding eigenvalues of the electronic Hamiltonian ${\hat C}_0$ are defined by:

\begin{equation}
{\hat C}_0 \left|n, \  l, \ m \right.\rangle = c_n \left|n, \  l, \ m \right.\rangle,
\end{equation}
with

\begin{equation}
c_n = -\frac{1}{n^2} \left(\frac{e^2}{4\pi \epsilon_0}\right)^2 \frac{m_e}{2\hbar^2} = -\frac{13.6 \ eV}{n^2},
\end{equation}
where $e$ is the elementary electric charge, $m_e$ is the reduced mass of the electron, $\epsilon_0$ is the permitivity of the vacuum and $\hbar$ is the reduced Planck constant. Thus, the eigenspace associated with the eigenvalue $c_n$ has linear dimension $\sum_{l=0}^{n-1} \sum_{m=-l}^{l} 1 = \sum_{l=0}^{n-1} \left(2 l +1\right) = n^2$, which is the degeneracy of this energy level.

Hence, according to the ABL rule, if the electron of the hydrogen atom is pre-selected in the state $|\Psi(t)\rangle$ and post-selected in the state $|\Phi(t)\rangle$, the probability to obtain the outcome $c_n$ in an hypothetical intermediate projective measurement at time $t$ is given by:

\begin{equation}
\label{conditional2}
p({\hat C}_0=c_n) = \frac{\left|\sum_{l=0}^{n-1}\sum_{m=-l}^{l} \langle \Phi(t)|n, l, m\rangle\langle n, l, m|\Psi(t)\rangle \right|^2}{\sum_{n'=1}^{\infty}  \left|\sum_{l'=0}^{n'-1}\sum_{m'=-l'}^{l'} \langle \Phi(t)|n', l', m'\rangle\langle n', l', m'|\Psi(t)\rangle \right|^2},
\end{equation}
which we can rewrite as:

\begin{equation}
\label{conditional3}
p({\hat C}_0=c_n) = \frac{\left|\sum_{m=-(n-1)}^{n-1}\sum_{l=|m|}^{n-1} \langle \Phi(t)|n, l, m\rangle\langle n, l, m|\Psi(t)\rangle \right|^2}{\sum_{n'=1}^{\infty}  \left|\sum_{m'=-(n'-1)}^{n'-1}\sum_{l'=|m'|}^{n'-1} \langle \Phi(t)|n', l', m'\rangle\langle n', l', m'|\Psi(t)\rangle \right|^2}.
\end{equation}
\\

On the other hand, in the presence of a weak external magnetic field $\epsilon B$ along the Z axis the electronic Hamiltonian gets the form ${\hat C} = {\hat C}_0 - \epsilon \frac{e}{2 m_e} B {\hat L}_z$, which partially breaks the degeneracy of the electronic energy levels:

 \begin{equation}
{\hat C} \left|n, \  l, \ m \right.\rangle = c_{n,m} \left|n, \  l, \ m \right.\rangle = \left(c_n - \epsilon \frac{e}{2 m_e} B \ m\right) \left|n, \  l, \ m \right.\rangle.
\end{equation}

Accordingly, the probability to obtain the outcome $c_{n, m}$ in a hypothetical intermediate projective measurement at time $t$ is given by:

\begin{equation}
\label{conditional4}
p({\hat C}=c_{n, m}) = \frac{\left|\sum_{l=|m|}^{n-1} \langle \Phi(t)|n, l, m\rangle\langle n, l, m|\Psi(t)\rangle \right|^2}{\sum_{n'=1}^{\infty}  \sum_{m'=-(n'-1)}^{n'-1} \left|\sum_{l'=|m'|}^{n'-1} \langle \Phi(t)|n', l', m'\rangle\langle n', l', m'|\Psi(t)\rangle \right|^2},
\end{equation}
which we can write as:

\begin{equation}
\label{conditional5}
p({\hat C}=c_{n, m}) = \frac{\left|\langle \Phi(t)|{\hat P}_{n, m}|\Psi(t)\rangle \right|^2}{\sum_{n'=1}^{\infty}  \sum_{m'=-(n'-1)}^{n'-1} \left|\langle \Phi(t)|{\hat P}_{n', m'}|\Psi(t)\rangle \right|^2},
\end{equation}
where

\begin{equation}
{\hat P}_{n,m} \equiv \sum_{l = |m|}^{n-1} \left|n, l, m\right.\rangle\langle\left.n, l, m\right|
\end{equation}
is the projector onto the eigenspace associated with the energy eigenvalue $c_{n,m}$.
\\

We can now define a generalized measurement $\left\{{\cal F}_{n,m}\right\}_{n \in \mathbb{N}, |m|<n}$ as follows:

\begin{equation}
\label{generalized}
{\cal F}_{n,m} =   \sum_{n',m'} \frac{{\cal Z}_{n',m',n,m}}{{\cal Z}_{n',m'}} \ {\hat P}_{n',m'},
\end{equation}
where

\begin{equation}
\sum_{n',m'} \equiv \sum_{n'=1}^{\infty} \sum_{m'=-(n'-1)}^{n'-1},
\end{equation}

\begin{equation}
{\cal Z}_{n',m',n,m} \equiv e^{-\frac{\left(c_{n',m'}-c_{n,m}\right)^2}{2\Delta^2}},
\end{equation}
and

\begin{equation}
{\cal Z}_{n',m'} \equiv \sqrt{\sum_{n,m} \left({\cal Z}_{n',m',n,m}\right)^2}.
\end{equation}
Hence, taking into account that:

\begin{equation}
P_{n,m} \cdot P_{n',m'} = \delta_{n,n'} \ \delta_{m,m'} \ P_{n,m},
\end{equation}
we find that:

\begin{equation}
{\cal F}^{\dagger}_{n,m}{\cal F}_{n,m} = \sum_{n',m'} \frac{\left({\cal Z}_{n',m',n,m}\right)^2}{\left({\cal Z}_{n',m'}\right)^2} \ {\hat P}_{n',m'},
\end{equation}
and

\begin{equation}
\sum_{n,m} {\cal F}^{\dagger}_{n,m}{\cal F}_{n,m} = \sum_{n',m'} {\hat P}_{n',m'} =  \hat{\mathbb{I}},
\end{equation}
as required. Moreover, it can be readily seen from (\ref{generalized}) that

\begin{equation}
{\cal F}_{n,m} \simeq \frac{1}{\sqrt{{\cal X}}} \cdot \sum_{(n',m') :  \left|c_{n',m'} - c_{n,m} \right| \simlt |\Delta} {\hat P}_{n', m'},
\end{equation}
where ${\cal X} \sim \mbox{Card}\left\{(n',m') :  \left|c_{n',m'} - c_{n,m} \right| \simlt |\Delta\right\}$ counts the number of energy levels whose distance to the level $c_{n,m}$ is smaller than the resolution of the measurement $\Delta$ and, therefore, cannot be actually distinguished from it, so that they should be taken as degenerate.

Only when $\Delta \rightarrow 0^+$ is much smaller from the energy gap between the level $c_{n,m}$ and all the others, this energy level can be completely resolved and we have:

\begin{equation}
{\cal F}_{n,m} \simeq {\hat P}_{n, m}.
\end{equation}

\section{Discussion}

Probabilities in quantum systems having both initial and final conditions are customarily calculated using the ABL formula \cite{ABL}, which describes the probabilities to obtain a certain eigenvalue of the measured operator at any intermediate time.

When this rule is blindly applied to systems with negligibly broken degeneracies, it apparently leads to a paradox - no matter how small the difference between the almost degenerated eigenvalues is, it could be detected through a finite jump in the probabilities of each of the possible outcomes (with respect to the limiting case in which the degeneracy is exactly restored).

The paradox is solved upon completing the physical scenario and taking into account the finite resolution of the measuring device, either by applying the ABL rule to the measured system and the measuring device as a whole or by describing the intermediate measurement performed on the system as a set of positive-operator valued measure elements.

It can then be explicitly shown that eigenvalues of the measured observable whose difference is much smaller than the resolution of the intermediate measurement must be treated as effectively degenerate if we wish to use the ABL rule for the measured system alone. In contrast, eigenvalues whose difference is larger than the resolution of the measurement device must be treated as non-degenerate. As a practical example, we presented a detailed analysis of conditional probabilities in a hydrogen atom where the rotational symmetry is partially broken by a weak external magnetic field through the Zeeman effect.

This analysis may support a general viewpoint according to which when addressing a physical problem one should first understand the governing physical principles, as well as the involved approximations and assumptions, before heading to the mathematical formalism. Otherwise, a misguided use of the formalism could lead to apparent paradoxes. In such cases one should return to the fundamental physical principles in order to trace the origin of the paradox and then solve it by employing a mathematical model which better accords with these principles. All this, of course, does not diminish the major importance of mathematical language in describing physical problems, but only calls for a careful use thereof which is aligned with the physics.

\begin{acknowledgments}

This work has been supported in part by the Israel Science Foundation Grant No. 1311/14.

\end{acknowledgments}

\end{document}